\documentclass[a4paper,11pt]{article}

\usepackage{jheppub}

\usepackage{graphicx}

\title{\boldmath Reanalysis of the BFKL Pomeron at the next-to-leading logarithmic accuracy}

\author{Xu-Chang Zheng ,}
\author{Xing-Gang Wu ,} \emailAdd{wuxg@cqu.edu.cn}
\author{Sheng-Quan Wang ,}
\author{Jian-Ming Shen,}
\author{Qiong-Lian Zhang}

\affiliation{Department of Physics, Chongqing University, Chongqing 401331, P.R. China}

\abstract{
We apply the principle of maximum conformality (PMC) to the Balitsky-Fadin-Kuraev-Lipatov (BFKL) Pomeron intercept at the next-to-leading logarithmic (NLL) accuracy. The PMC eliminates the conventional renormalization scale ambiguity by absorbing the non-conformal $\{\beta_i\}$-terms into the running coupling, and a more accurate pQCD estimation can be obtained. After PMC scale setting, the QCD perturbative convergence can be greatly improved due to the elimination of renormalon terms in pQCD series, and the BFKL Pomeron intercept has a weak dependence on the virtuality of the reggeized gluon. For example, by taking the Fried-Yennie gauge, we obtain $\omega_{\rm MOM}^{\rm PMC}(Q^{2},0)\in [0.149,0.176]$ for $Q^2\in[1,100]\;{\rm GeV}^2$. This is a good property to apply to the high-energy phenomenology. Further more, to compare with the data, it is found that the physical ${\rm MOM}$-scheme is more reliable than the $\overline{\rm MS}$-scheme. The ${\rm MOM}$-scheme is gauge dependent, which can also be greatly suppressed after PMC scale setting. We discuss the MOM-scheme gauge dependence for the Pomeron intercept by adopting three gauges, i.e. the Landau gauge, the Feynman gauge and the Fried-Yennie gauge, and we obtain $\omega_{\rm MOM}^{\rm PMC}(Q^{2}=15\;{\rm GeV}^2,0) = 0.166^{+0.010}_{-0.017}$; i.e. about $10\%$ gauge dependence is observed. We apply the BFKL Pomeron intercept to the photon-photon collision process, and compare the theoretical predictions with the data from the OPAL and L3 experiments. \\

\noindent PACS number(s): 12.40.Nn, 12.38.Aw, 12.38.Cy }


\arxivnumber{1308.2381}

\begin{document}

\maketitle
\flushbottom

\section{Introduction}
\label{sec:1}

The high energy limit of QCD is one of the most important aspects of strong interactions. There are always more than one energy scale at high-energy hadronic collisions. If the ratios of these scales are very large, or there are large log terms involving a particular scale, we should resum these large logarithms so as to achieve a reliable estimation. For example, the deep inelastic scattering at small Bjorken $x$ involves large logarithm as $\ln s$, with $\sqrt{s}$ the $e^+ e^-$ collision energy. A resummation of $[\alpha_{s}\ln s]^{n}$ is displayed by the leading logarithmic (LL) Balitsky-Fadin-Kuraev-Lipatov (BFKL) equation \cite{LLBFKL1,LLBFKL2,LLBFKL3,LLBFKL4}. The Pomeron, built from gluons, is a coherent color-singlet object, and in LL approximation, it is treated as a compound state of two reggeized gluons. The maximum eigenvalue ($\omega^{max}$) of the BFKL equation relates to the BFKL Pomeron intercept, which dominates the high-energy asymptotic behavior of the scattering cross section via the relation, $\sigma\varpropto s^{\alpha_{P}-1}=s^{\omega^{max}}$. Thus it is interesting and necessary to study the properties of the Pomeron intercept. As has already been shown that the leading order (LO) $\omega^{max}$ is about 0.55 for $\alpha_{s}=0.2$, this is rather large compared to experiments. So, the next-to-leading order (NLO) correction to the BFKL equation is very important, which can be obtained by the resummation of $\alpha_{s}[\alpha_{s}\ln s]^{n}$ terms. One may expect to achieve a reasonable intercept with the higher-order corrections at the next-to-leading logarithmic (NLL) accuracy~\cite{NLLBFKL1,NLLBFKL2,NLLBFKL3,NLLBFKL4,NLLBFKL5,NLLBFKL6,add2,add6,add7}. The Pomeron intercept has also been analyzed in ${\cal N}=4$ super symmetry theory~\cite{add3,add4}. It is shown that in certain large-${\cal N}$ QCD-like theories, the BFKL regime and the classic soft Regge regime can be simultaneously described by using the curved-space string theory.

For calculating higher-order contributions to the Pomeron intercept, one needs to introduce the renormalization scale $\mu_R$. In the literature, it is usually taken as the typical momentum flow of the process ($Q$), which can eliminate the (possible large) log terms involving $\ln (\mu_R/Q)$ to a certain degree. However for the non-Abelian QCD case, it is not so simple and the naive choice of $Q$ shall lead to unreasonable large NLO corrections, cf. Refs.~\cite{NLLBFKL1,NLLBFKL2}. Under such conventional scale setting, it has been found that the Pomeron intercept includes both the renormaliztion scheme and the renormalization scale ambiguities. Such renormalization scale uncertainty usually provides a large systematic error under the conventional scale setting. To cure the scale ambiguity, in Ref.\cite{Pomeron}, the authors have suggested to use Brodsky-Lepage-Mackenzie (BLM) scale setting~\cite{BLM} to deal with the Pomeron intercept. After applying BLM scale setting and by transforming the expressions under the modified minimal subtraction scheme ($\overline{\rm MS}$)~\cite{MSbar} into those under the momentum space subtraction scheme (MOM)~\cite{MOM}, they obtained the Pomeron intercept $\omega^{max}=0.13\thicksim0.18$. The scheme transformation from $\overline{\rm MS}$ to MOM is necessary, since as they have found that the NLO terms give unreasonablely large contributions to the Pomeron intercept under the $\overline{\rm MS}$ scheme; while by using the MOM-scheme, a reasonable intercept can be obtained.

Recently, it has been pointed out that the principle of maximum conformality (PMC) provides a possible systematic solution for eliminating the renormalization scheme and renormalization scale ambiguities~\cite{PMC1,PMC2,PMC3,PMC4,PMC5,PMC6,PMC7,PMC8,PMC9}. The PMC provides the underlying principle for BLM scale setting~\cite{BLM}, and it satisfies all the necessary self-consistency requirements of the renormalization group~\cite{PMC5}. As a step forward of Ref.\cite{Pomeron}, in this paper, we shall apply PMC scale setting to deal with the NLL BFKL Pomeron intercept.

The main idea of PMC lies in that one can first finish the renormalization procedure for any pQCD process by using an arbitrary renormalization scheme and an arbitrary initial renormalization scale ($\mu_R^{\rm init}$), and then set the effective or optimal PMC scale for the process. In comparison, under conventional scale setting, one always fixes the renormalization scale to be $\mu_R^{\rm init}$. The PMC scale ($\mu_R^{\rm PMC}$) is formed by absorbing all non-conformal terms that governs the running behavior of the coupling constant into the coupling constant. At each perturbative order, new $\{\beta_i\}$-terms will occur, so the PMC scale for each perturbative order is generally different. To be consistent, similar to the case of BLM scale setting~\cite{comm1,comm2,comm3}, the PMC scales themselves are also in perturbative series. This property have been put in a more solid background by using the newly suggested $R_\delta$-scheme~\cite{PMC7,PMC8}, which provides an elegant way to demonstrate the PMC principle and a method to automatically setting the PMC scales to all-orders. Even though, one may choose any arbitrary value to be $\mu^{\rm init}_R$, the optimal PMC scales and the resulting finite-order PMC prediction are both to high accuracy independent of such arbitrariness, consistent with the renormalization group invariance. There is residual initial renormalization-scale dependence due to the lack of information on even higher-order $\{\beta_i\}$-terms. Such residual scale-uncertainty will be greatly suppressed when the PMC scales have been set suitably. After PMC scale setting, the divergent ``renormalon" series $(n!\;\beta_i^{n}\alpha_s^n)$ does not appear and the convergence of the pQCD series can be greatly improved in principle.

The remaining parts of this paper are organized as follows. In Sec.\ref{sec:2}, we present the calculation technology for dealing with the eigenvalue of the BFKL equation and the Pomeron intercept at the NLL level by applying PMC scale setting. In Sec.\ref{sec:3}, we give a detailed discussion on the properties of the Pomeron intercept and give a comparison of the photon-photon collision process between our estimates and the OPAL and L3 data~\cite{OPAL,L3}. Sec.\ref{sec:4} is reserved for a summary.

\section{The PMC scale setting for the eigenvalue of the BFKL equation and the Pomeron intercept} \label{sec:2}

\subsection{The eigenvalue of the BFKL equation and the Pomeron intercept}

Schematically, the eigenvalue of the BFKL equation $\omega_{\overline{\rm MS}}(Q^{2},\nu)$ under the $\overline{\rm MS}$ scheme, which up to NLL level can be expressed as
\begin{eqnarray}
\omega_{\overline{\rm MS}}(Q^{2},\nu) =N_{C}~\chi_L(\nu) \frac{\alpha_{\overline{\rm MS}}(\mu^{\rm init}_R)}{\pi}\left[1+r_{\overline{\rm MS}}(Q,\mu^{\rm init}_R,\nu) \frac{\alpha_{\overline{\rm MS}}(\mu^{\rm init}_R)}{\pi} + {\cal O}\left( \left( \frac{\alpha_{\overline{\rm MS}}}{\pi} \right)^2 \right) \right],
\label{omegams1}
\end{eqnarray}
where $Q$ stands for the virtuality of the reggeized gluon, $\nu$ is the conformal weight parameter, $N_{C}(=3)$ is the number of colors. It is an (good) approximation that $\omega_{\overline{\rm MS}}(Q^{2},\nu)$ stands for the eigenvalue of the BFKL equation, since it is obtained by averaging the BFKL kernel over the azimuthal angles and the multi-gluon components of the Pomeron wavefunction are neglected~\cite{NLLBFKL1,NLLBFKL2}. The intercept of the BFKL Pomeron is related to the maximum eigenvalue of the BFKL equation $\omega^{max}_{\overline{\rm MS}}$, which can be obtained by taking the limit $\nu\to 0$~\cite{linv}. Here we have used $\mu^{\rm init}_R$ to stand for the initial/formal renormalization scale, which is usually taken as the typical momentum flow of the process (or at which the experiment is performed) that can absorb all the indeterminate radiative corrections into the definition of the coupling constant \footnote{In fact, such argument is conceptional and is not strict, since it is the $\{\beta_i\}$-functions that rightly governs the correct behavior of the coupling constant.}. Substituting the conventional choice of $\mu_R \equiv \mu^{\rm init}_R = Q$ into Eq.(\ref{omegams1}), we return to the usual expression for $\omega_{\overline{\rm MS}}(Q^{2},\nu)$ as shown in Ref.\cite{NLLBFKL1}. Under such conventional scale setting, the renormalization scale uncertainty usually provides a large systematic error for pQCD estimations, which can be roughly estimated by varying the scale within the region of $[Q/2,2Q]$. In the present paper, we shall apply PMC to deal with the process and to show how the scale uncertainty can be greatly suppressed. Thus our understanding of the Pomeron properties can be greatly improved. For the purpose, we keep $\mu^{\rm init}_R$ as a free parameter, which may or may not equal to $Q$, then the terms proportional to $\ln\left[ {Q^2}/{\left(\mu_R^{\rm init}\right)^2} \right]$ should be kept.

For convenience of applying PMC scale setting, we further decompose the NLO coefficient $r_{\overline{\rm MS}}(Q,\mu^{\rm init}_R,\nu)$ into two parts,
\begin{eqnarray}
r_{\overline{\rm MS}}(Q,\mu^{\rm init}_R,\nu)&=& r_{\overline{\rm MS}}^{\rm conf}(Q,\mu^{\rm init}_R,\nu) + r_{\overline{\rm MS}}^{\beta}(Q,\mu^{\rm init}_R,\nu) \; \beta_0,
\end{eqnarray}
with the coefficient of the non-conformal part,
\begin{eqnarray}
r_{\overline{\rm MS}}^{\beta}(Q,\mu^{\rm init}_R,\nu)&=&-\frac{1}{4} \left[\frac{1}{2}\chi_L(\nu)+ \ln\left(\frac{Q^2}{\left(\mu_R^{\rm init}\right)^2} \right)-\frac{5}{3}\right]
\label{rms0:nu}
\end{eqnarray}
and the coefficient of the conformal part
\begin{eqnarray}
r_{\overline{\rm MS}}^{\rm conf}(Q,\mu^{\rm init}_R,\nu) &=&-\frac{N_C}{4\chi_L(\nu)} \left[\frac{\pi^2 \rm sinh(\pi \nu)}{2 \nu~\rm cosh^2(\pi \nu)}\left(3+\left(1+\frac{n_f}{N_C^3}\right)\frac{11+12\nu^2} {16(1+\nu^2)}\right)-\chi_L^{''}(\nu) \right. \nonumber \\
&& \left. \quad\quad\quad\quad\quad  +\frac{\pi^2-4}{3}\chi_L(\nu)-\frac{\pi^3}{\rm cosh(\pi \nu)}- 6 \zeta(3)+4\varphi(\nu)\right]. \label{rms0:nu:conf}
\end{eqnarray}
Here,
\begin{eqnarray}
\varphi(\nu)=2\int_0^1dx \frac{\rm cos(\nu~ln~x)}{(1+x)\sqrt{x}}\left[\frac{\pi^2}{6}-\rm Li_2(x)\right],
\rm Li_2(x)=-\int_0^x dt \frac{\rm ln(1-t)}{t}
\end{eqnarray}
and
\begin{eqnarray}
\chi_L(\nu)=2\psi(1)-\psi(1/2+i \nu)-\psi(1/2-i \nu).
\end{eqnarray}
The $\psi(\gamma)=\Gamma^{'}(\gamma)/\Gamma(\gamma)$ is Euler $\psi$-function. Note that in higher-order processes, there are $n_f$-terms coming from the Feynman diagrams with the light-by-light quark loops which are irrelevant to the ultra-violet cutoff. In Eq.(\ref{rms0:nu:conf}), the $n_f$ stands for the number of quark flavors from the Abelian part of the process $gg\rightarrow q\overline{q}$. Those $n_f$-terms have no relation to the $\{\beta_i\}$-terms, and they should be identified and kept separate from the PMC scale setting~\cite{PMC9}. So we set this $n_f=4$ through this paper. This treatment agrees with the observation of Ref.\cite{Binger}, in which it shows that twelve of the thirteen invariant amplitudes due to light-quark loop contributions to the three-gluon coupling give an $n_f$ dependence in pQCD; however, this $n_f$ dependence is not associated with the QCD $\{\beta_i\}$-terms and should be kept separate during the scale setting.

\subsection{The PMC scale setting}

Before applying the PMC procedures to the process, we need to transform the results from the $\overline{\rm MS}$-scheme to those of a non-Abelian physical scheme such as the MOM-scheme. This is because that for the processes involving three-gluon vertex, the scale setting problem is much more involved. It has already been observed that the renormalization scale which appears in the three-gluon vertex should be a function of the virtuality of the three external gluons $q^2_1$, $q^2_2$ and $q^2_3$~\cite{Binger}. More explicitly, the analysis in Ref.\cite{Binger} shows that when the virtualities are very different as in the subprocess $g g \to g \to Q \bar Q$, the squared renormalization scale $\mu^{2}_R$ should be proportional to ${q^2_{\rm min}  q^2_{\rm med} \over q^2_{\rm max}}$, where $|q^2_{\rm min}| < |q^2_{\rm med}| < |q^2_{\rm max}|$ with $q^2_{\rm max}$ being the maximal virtuality. A naive prediction based on the guessing scale $\mu^2_R \simeq Q^2$ will give misleading result, especially for the present BFKL case in which there are important leading-order gluon-gluon interactions. This could be the reason why under the $\overline{\rm MS}$-scheme, the NLO correction to the maximum eigenvalue is negative and even larger than LO contribution for $\alpha_{s}>0.157$~\cite{NLLBFKL1,NLLBFKL2}.

On the other hand, the non-Abelian physical scheme, e.g. the MOM scheme, based on the renormalization of the symmetric triple-gluon vertex is appropriate for the purpose~\cite{MOM}. The MOM scheme is implemented by prescribing the values of divergent propagators and vertices at some fixed configuration of external momenta, which is consistent with the treatment of Ref.\cite{Binger}. Due to the commensurate scale relation among different renormalization, the physical estimation is indifferent to various scheme choices. Whereas, for a fixed-order calculation, the pQCD series after PMC scale setting is only nearly conformal due to unknown $\{\beta_i\}$-terms, the pQCD convergence could be different for different choice of schemes~\cite{PMC9}. Because the MOM scheme is a physical method of renormalization, one may expect reasonable convergence from expansions of physical quantities in terms of MOM coupling constant.

The transformation from the $\overline{\rm MS}$ scheme to the MOM scheme can be accomplished by a transformation of the QCD coupling constant. At the one loop level, we have~\cite{MOM}:
\begin{eqnarray}
\alpha_{\overline{\rm MS}} = \alpha_{\rm MOM}\left[ 1+T_{\overline{\rm MS}/{\rm MOM}}(\xi) \frac{\alpha_{\rm MOM}}{\pi} + {\cal O}\left(\left(\frac{\alpha_{\rm MOM}}{\pi}\right)^{2} \right) \right] ,  \label{schemetransformation}
\end{eqnarray}
where the gauge dependent $T$-function can also be decomposed into the $\beta$-dependent and $\beta$-independent parts
\begin{equation}
T_{\overline{\rm MS}/{\rm MOM}}(\xi)= T_{\overline{\rm MS}/{\rm MOM}}^{\rm conf}(\xi) + T_{\overline{\rm MS}/{\rm MOM}}^\beta(\xi) \; \beta_0  \label{msmom}
\end{equation}
with
\begin{equation}
T_{\overline{\rm MS}/{\rm MOM}}^{\rm conf}(\xi) =\frac{N_c}{8} \left[\frac{17}{2}I +\frac{3}{2}(I-1)\xi+\left(1-\frac{1}{3}I\right) \xi^2 -\frac{1}{6}\xi^3\right]
\end{equation}
and
\begin{equation}
T_{\overline{\rm MS}/{\rm MOM}}^\beta(\xi)=-\frac{1}{2}\left(1+\frac{2}{3}I\right) ,
\end{equation}
where $I=\int_0^1 [{\ln x^2}/{(x-x^2-1)}] dx\simeq2.344$ and $\xi$ is the gauge parameter for MOM scheme. The $\beta$-dependent $T$-function ($T_{\overline{\rm MS}/{\rm MOM}}^\beta$) together with the coefficient ($r_{\overline{\rm MS}}^{\beta}$) rightly governs the running behavior of the coupling constant, it should be absorbed into the coupling constant. It is noted that the MOM scheme is gauge dependent already at the leading-order level, e.g. $\xi=0$ stands for the Landau gauge, $\xi=1$ stands for the Feynman gauge and $\xi=3$ stands for the Fried-Yennie gauge. The Fried-Yennie is helpful for the bound state problems, which can alleviate the notorious infrared difficulties in such kind of problems~\cite{yennie,yennie2}. In the following we shall show how the gauge parameter affects the final estimation.

As a combination of Eqs.(\ref{omegams1},\ref{schemetransformation}), we obtain the eigenvalue of the BFKL equation $\omega_{\rm MOM}(Q^{2},\nu)$ under the ${\rm MOM}$ scheme at the NLL level,
\begin{displaymath}
\omega_{\rm MOM}(Q^{2},\nu) = N_C~\chi_L(\nu) \frac{\alpha_{\rm MOM}(\mu^{\rm init}_R)}{\pi} \left[1+ \left(r_{\overline{\rm MS}}(Q,\mu^{\rm init}_R,\nu) +T_{\overline{\rm MS}/{\rm MOM}}(\xi)\right) \frac{\alpha_{\rm MOM}(\mu^{\rm init}_R)}{\pi}\right] .
\end{displaymath}
After applying PMC scale setting, we obtain the NLL BFKL eigenvalue $\omega_{\rm MOM}^{\rm PMC}(Q^{2},\nu)$,
\begin{eqnarray}
\omega_{\rm MOM}^{\rm PMC}(Q^{2},\nu) &=& N_C~\chi_L(\nu) \frac{\alpha_{\rm MOM}(\mu_R^{\rm PMC})} {\pi} \left[1+ C^{\rm conf}(Q,\mu^{\rm init}_R,\nu,\xi) \frac{\alpha_{\rm MOM}(\mu_R^{\rm PMC})}{\pi}\right]
\end{eqnarray}
with the conformal term $$C^{\rm conf}(Q,\mu^{\rm init}_R,\nu,\xi)= \left(r_{\overline{\rm MS}}^{\rm conf}(Q,\mu^{\rm init}_R,\nu) +T_{\overline{\rm MS}/ {\rm MOM}}^{\rm conf}(\xi)\right) $$ and the LO PMC scale, which eliminates the non-conformal $\beta_0$-term of the perturbative series, is
\begin{eqnarray}
\mu_R^{\rm PMC}= Q~{\rm exp} \left[\frac{1}{4}\chi_L(\nu) -\frac{5}{6} +\left(1+\frac{2}{3}I \right)\right] , \label{scale:MOM:PMC}
\end{eqnarray}
which is explicitly independent of the initial scale $\mu^{\rm init}_R$. This shows that the PMC scale and the $\omega_{\rm MOM}^{\rm PMC}(Q^{2},\nu)$ are initial scale independent, and hence the usual renormalization scale dependence is accidentally eliminated at the NLO level.

There are two types of residual scale dependence for the present NLO estimation due to unknown higher order $\{\beta_i\}$-terms: 1) The residual scale dependence for the LO PMC scale $\mu_R^{\rm PMC;LO}$, which is highly exponentially suppressed, i.e. those unknown $\{\beta_i\}$-terms shall be absorbed into the perturbative part of the PMC scale $\mu_R^{\rm PMC;LO}$ itself as an exponential factor~\cite{PMC9}; 2) We have set the NLO PMC scale to be equal to the LO PMC scale because of lacking NNLO $\{\beta_i\}$-terms, i.e. $\mu_R^{\rm PMC;NLO}=\mu_R^{\rm PMC;LO}=\mu_R^{\rm PMC}$. Roughly, similar to the conventional scale setting, one can estimate the uncertainty of NLO terms by varying $\mu_R^{\rm PMC;NLO}\in \left[\mu_R^{\rm PMC;LO}/2, 2\mu_R^{\rm PMC;LO}\right]$. The examples for the small residual (initial) scale dependence at higher orders, e.g. the two-loop top pair production and the four-loop $R(e^+ e^-)$, can be found in Refs.\cite{PMC1,PMC2,PMC3,PMC4,PMC7,PMC8,PMC9}.

Since the intercept of the BFKL Pomeron is related to the maximum eigenvalue of the BFKL equation $\omega_{\rm MOM}^{\rm PMC}(Q^{2},\nu)$, we can conveniently obtain the final results for the Pomeron intercept by taking the limit $\nu\to0$ in the above expressions.

\subsection{A more accurate estimation from the extended renormalization group}

If two renormalization schemes are quite different, then the fixed-order coupling constant transformation as Eq.(\ref{schemetransformation}) may not be good enough to guarantee a well pQCD convergence. That is, we need an optimal scale setting to get the estimation as accurate as possible by using the known fixed-order results. For the purpose, the extended renormalization group~\cite{PMC1,extended,HJLu} that transforms the schemes also in a similar continuous way as that of the scales can be adopted for a more reliable estimation.

More explicitly, as an extension of the ordinary coupling, one can define a universal coupling $a(\tau,\{c_i\})={\beta_1}/{(4\pi\beta_0)} \alpha$ with $c_i = {\beta_i \beta_0^{i-1}} / {\beta^i_1}$ to include its dependence on both the scale parameter $\tau={\beta^2_0}/{\beta_1} \ln\left(\mu^2_R/\Lambda^2_{\rm QCD}\right)$ and the scheme parameters $\{c_i\}$. As usual, the scale evolution equation for the universal coupling can be written as
\begin{equation}
\beta(a,\{c_i\}) = \frac{\partial a}{\partial \tau} = -a^2 \left[1+ a +c_2 a^2+c_3 a^3 +\cdots \right] . \label{scale}
\end{equation}
and the scheme evolution equation for the universal coupling can be defined as
\begin{equation}
\beta_n(a(\tau,\{c_i\}),\{c_i\}) = \frac{\partial}{\partial c_n} a(\tau,\{c_i\}) ,
\end{equation}
which leads to
\begin{equation}
\beta_n(a(\tau,\{c_i\}),\{c_i\}) = -\beta(a(\tau,\{c_i\}),\{c_i\}) \int_0^{a(\tau,\{c_i\})} \frac{ x^{n+2} dx}{\beta^2(x,\{c_i\})}, \label{scheme}
\end{equation}
where $a(0,\{0\})=\infty$ and $\beta(0,\{0\})=0$ are boundary conditions, and the lower limit of the integral has been set to satisfy the boundary condition $\beta_{n}(a(\tau,\{c_i\}),\{c_i\})={\cal O}(a^{n+1})$. This means that the truncation of the $\{\beta_i\}$-functions simply corresponds to evaluating $a(\tau,\{c_i\})$ in a subspace where higher-order $c_i$ are zero. The scheme-equation (\ref{scheme}) can be used to relate the couplings under different schemes by properly changing $\{c_i\}$.

It is noted that Eq.(\ref{scheme}) can be perturbatively solved with the help of the scale-equation (\ref{scale}), which can be adopted to estimate how the uncalculated higher-order terms contribute to the final result. In the following we show that a more strict relation between the coupling constant under the $\overline{\rm MS}$-scheme and the MOM-scheme can be obtained from the extended renormalization group.

As a first step, we write down the relation between the QCD coupling constant under the $\overline{\rm MS}$-scheme and the MOM-scheme up to NNLO level~\cite{MOM}:
\begin{eqnarray}
\alpha_{\overline{\rm MS}} = \alpha_{\rm MOM}\left[ 1 + T_{1} \frac{\alpha_{\rm MOM}}{\pi} + T_{2} \left(\frac{\alpha_{\rm MOM}}{\pi}\right)^{2} + {\cal O}\left(\left(\frac{\alpha_{\rm MOM}}{\pi}\right)^{3} \right)\right] ,\label{schemetransformation2}
\end{eqnarray}
where under the Landau gauge, the NLO coefficient $T_{1}$ equals to $T_{\overline{\rm MS}/{\rm MOM}}(0)$ that is defined in Eq.(\ref{msmom}) and the NNLO coefficient
\begin{eqnarray}
T_{2}=\left(2d_{10}^2-d_{20}\right) + \left( 4d_{10}d_{11}-d_{21}\right)n_f + \left(2d_{11}^2-d_{22}\right)n_f^2 ,
\end{eqnarray}
where we have~\cite{MOM2}: $d_{10}=11/2+23 I/48$, $d_{11}=-1/3-2 I/9$, $d_{20}=59.8$, $d_{21}=-12.6$ and $d_{22}=47/432+7 I/54 +  I^2/81$. Those $n_f$ series can be written as the $\{\beta_i\}$-series via the PMC-BLM correspondence principle suggested in Ref.\cite{PMC1}, or the  newly suggested equivalent method presented in Refs.\cite{PMC7,PMC8}.

Further more, we adopt the extended renormalization group equation to get a more accurate relation between the MOM-scheme and the $\overline{\rm MS}$-scheme. Using the redefinition of $a={\beta_1}/{(4\pi\beta_0)} \alpha$, we rewrite Eq.(\ref{schemetransformation2}) in the following,
\begin{eqnarray}
a_{\overline{\rm MS}} &=& a_{\rm MOM}\left[ 1+\left(\frac{4 \beta_0}{\beta_1}\right) T_{1}\; a_{\rm MOM} + \left(\frac{4 \beta_0} {\beta_1}\right)^2 T_{2}\; a_{\rm MOM}^{2}+{\cal O}\left(a_{\rm MOM}^{3}\right) \right] \label{schemetransformation3} \\
&=&  a_{\rm MOM}\left[ 1+f_2 \; a_{\rm MOM} + f_3 \; a_{\rm MOM}^{2}+{\cal O}\left(a_{\rm MOM}^{3}\right) \right].
\end{eqnarray}
With the help of the extended renormalization equations, we obtain~\cite{HJLu}
\begin{equation}
\tau_{\overline{\rm MS}}=\tau_{{\rm MOM}}-f_2
\end{equation}
\begin{equation}
c_2^{\overline{\rm MS}}=c_2^{{\rm MOM}} - f_2 - f_2^2 + f_3,
\end{equation}
Here the first equation gives the relation between the asymptotic scale under the MOM-scheme and the $\overline{\rm MS}$-scheme~\cite{MOM}, i.e. ${\Lambda^{\rm MOM}_{\rm QCD}} /{\Lambda^{\overline{\rm MS}}_{\rm QCD}} = \exp[-2T_{1}/\beta_0]$. Because the second equation is derived from the coupling constant under the $\overline{\rm MS}$-scheme through a continuous transformation, then, we can use the renormalization scale equation (\ref{scale}) to derive a more accurate running behavior for the coupling constant at the NNLO level. Such a behavior shall be helpful if we have known the NNLO eigenvalue of the BFKL equation and hence the NNLO Pomeron intercept in the future.

\section{Numerical results and discussions}
\label{sec:3}

For numerical calculation, we adopt the two-loop running coupling together with the reference point $\alpha_{\overline{\rm MS}}(m_Z)=0.1184$~\cite{pdg} to set the asymptotic scale, i.e.
\begin{displaymath}
\Lambda^{\overline{\rm MS}}_{\rm QCD}|_{(n_f=3)}=0.386 \; {\rm GeV},\; \Lambda^{\overline{\rm MS}}_{\rm QCD}|_{(n_f=4)}=0.332\; {\rm GeV}\;\; {\rm and}\;\; \Lambda^{\overline{\rm MS}}_{\rm QCD}|_{(n_f=5)}=0.231\; {\rm GeV}.
\end{displaymath}
Using the relation between the asymptotic scale, ${\Lambda^{\rm MOM}_{\rm QCD}}/{\Lambda^{\overline{\rm MS}}_{\rm QCD}} = \exp[-2T_{\overline{\rm MS}/{\rm MOM}}(\xi)/\beta_0]$, one can conveniently obtain the asymptotic scale under the MOM scheme. For example, under the Landau gauge, we have
\begin{displaymath}
\Lambda^{\rm MOM}_{\rm QCD}|_{(n_f=3)}=0.952\; {\rm GeV}, \; \Lambda^{\rm MOM}_{\rm QCD}|_{(n_f=4)}=0.718\; {\rm GeV}\;\; {\rm and} \;\; \Lambda^{\rm MOM}_{\rm QCD}|_{(n_f=5)}=0.427\; {\rm GeV}.
\end{displaymath}

\subsection{Properties of the NLO BFKL Pomeron intercept}

\begin{figure}[htb]
\centering
\includegraphics[width=0.49\textwidth]{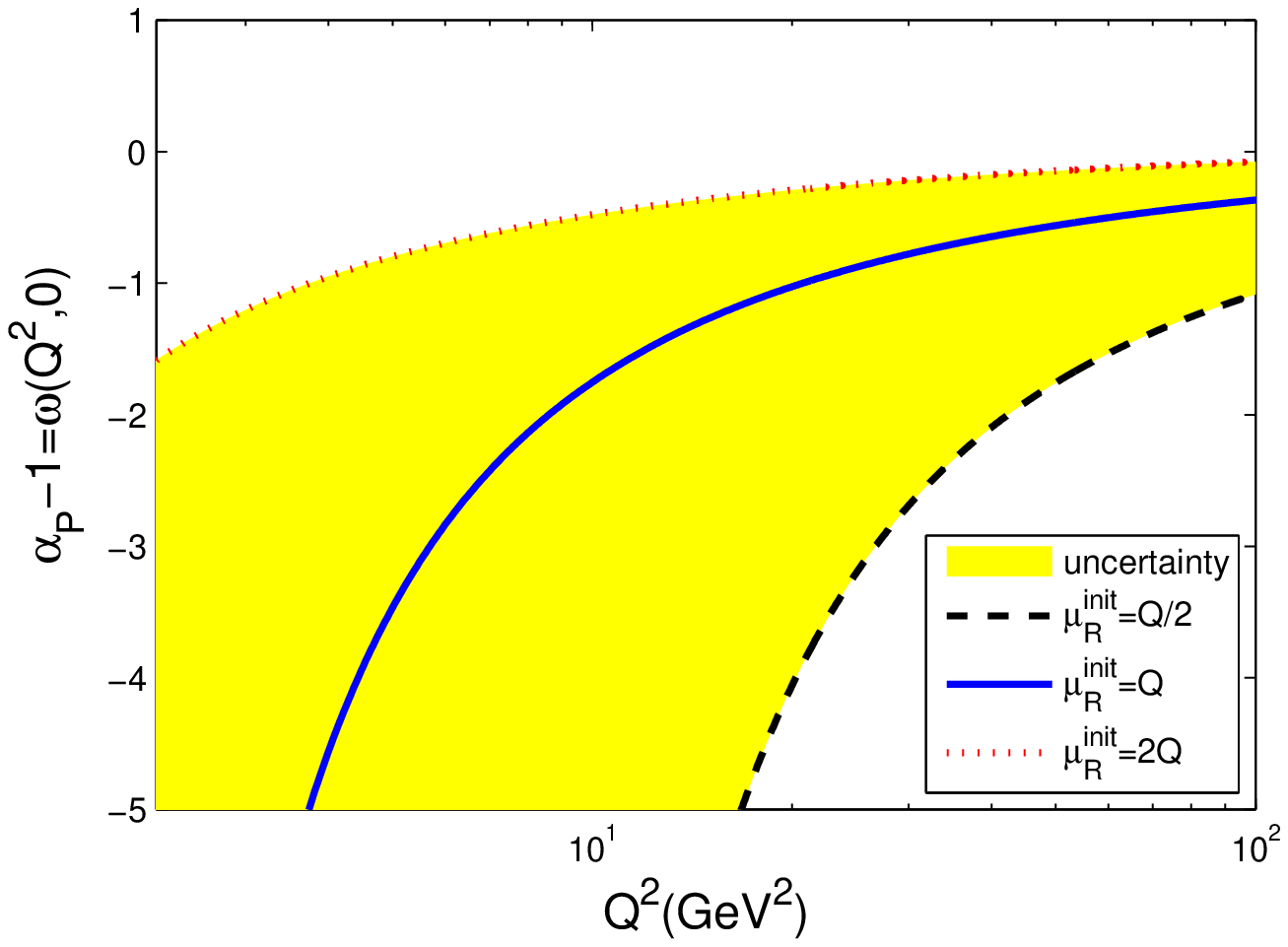}
\includegraphics[width=0.49\textwidth]{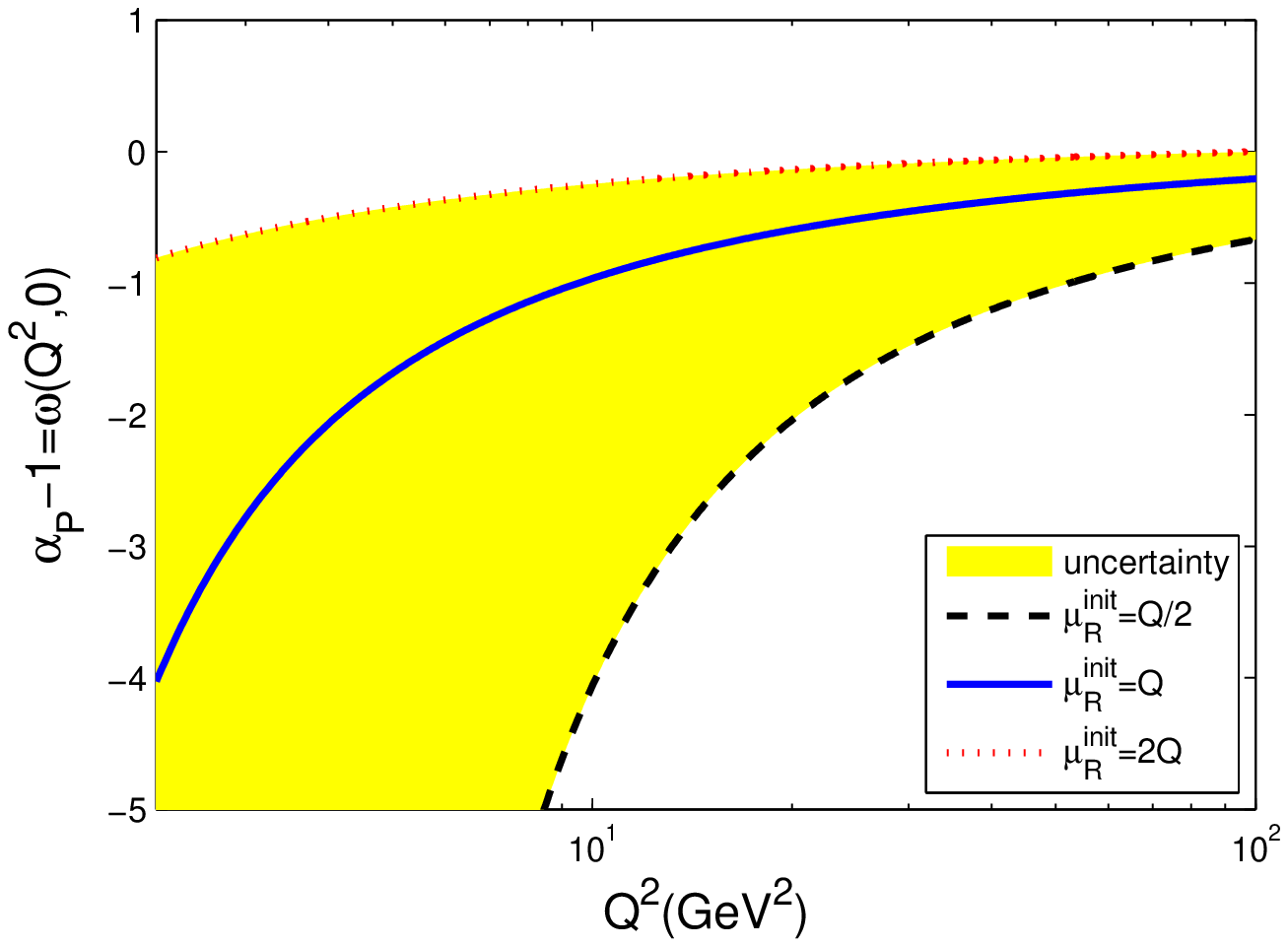}
\caption{The error analysis of the NLO BFKL Pomeron intercept $\omega(Q^{2},0)$ versus $Q^2$ under the ${\rm MOM}$ scheme and the conventional scale setting. The shaded band shows the conventional renormalization scale uncertainty by varying the renormalization scale $\mu_R (\equiv\mu_R^{\rm init})$ within the region of $[Q/2, 2Q]$. The left diagram is for the Landau gauge with $\xi=0$ and the right one is for the Fried-Yennie gauge with $\xi=3$. } \label{disscale}
\end{figure}

The NLO BFKL Pomeron intercept $\omega(Q^{2},0)$ is obtained by taking the limit $\nu\to0$ to the eigenvalue of the BFKL equation. As a comparison, we firstly present the NLO BFKL Pomeron intercept $\omega(Q^{2},0)$ under the ${\rm MOM}$ scheme and the conventional scale setting in Fig.(\ref{disscale}). Here as usual, the scale uncertainty is estimated by taking $\mu_R\equiv\mu^{\rm init}_{R} \in [Q/2,2Q]$. Fig.(\ref{disscale}) shows

\begin{itemize}
\item The NLO Pomeron intercept relies heavily on the virtuality of the reggeized gluon, its magnitude shows a fast increase with the increment of $Q^2$, especially for the case of $\mu_R=Q/2$. This means that the log-terms proportional to $\ln\mu_R/Q$ eliminated by the choice of $\mu_R=Q$ shall have large contributions for other scale choices, and we need to know NNLO terms in order to provide a more reliable estimation.

    Recently, it has been suggested that by following the idea of PMC, one can achieve a better scale-error estimation under the conventional scale setting by partly including the coupling constant behaviors at higher orders into the prediction~\cite{pmcnew1}. Such an approximate analysis can be achieved by expanding the coupling constant at any scale $\mu_R$ to be around $Q$, i.e.,
    \begin{equation}
     a_{s}(\mu_R) = a_{s}(Q)-\beta_{0} \ln\left(\frac{\mu_R^{2}}{Q^2}\right) a^{2}_{s}(Q) +\left[\beta^2_{0} \ln^2 \left(\frac{\mu_R^{2}}{Q^2}\right) -\beta_{1} \ln\left(\frac{\mu_R^{2}}{Q^2}\right) \right] a^{3}_{s}(Q) + {\cal O}(a^{4}_{s}) , \label{alphasrun}
     \end{equation}
    where $a_s=\alpha_s/4\pi$. These retrieved log-terms for $\mu_R\neq Q$ at higher orders which can largely compensate the scale changes at the NLO level and then result in a more reasonable conventional scale error estimation.

\item Under conventional scale setting, there is large scale uncertainty, and similar to the case of $\overline{\rm MS}$ scheme~\cite{NLLBFKL1,NLLBFKL2}, we also obtain a negative value for $\omega(Q^{2},0)$ under the ${\rm MOM}$ scheme. The large negative NLO contribution shows the pQCD convergence is questionable. Thus, we need to know the more complex NNLO or even higher-order correction so as to obtain a reliable pQCD estimation.

\item There is gauge dependence under the MOM scheme, e.g. different choice of $\xi$ shall lead to different estimations. Such gauge dependence also heavily depends on the choice of renormalization scale.
\end{itemize}

After applying PMC scale setting, we show that the theoretical estimation on the NLO BFKL Pomeron intercept can be greatly improved :
\begin{enumerate}

\begin{figure}
\centering
\includegraphics[width=0.49\textwidth]{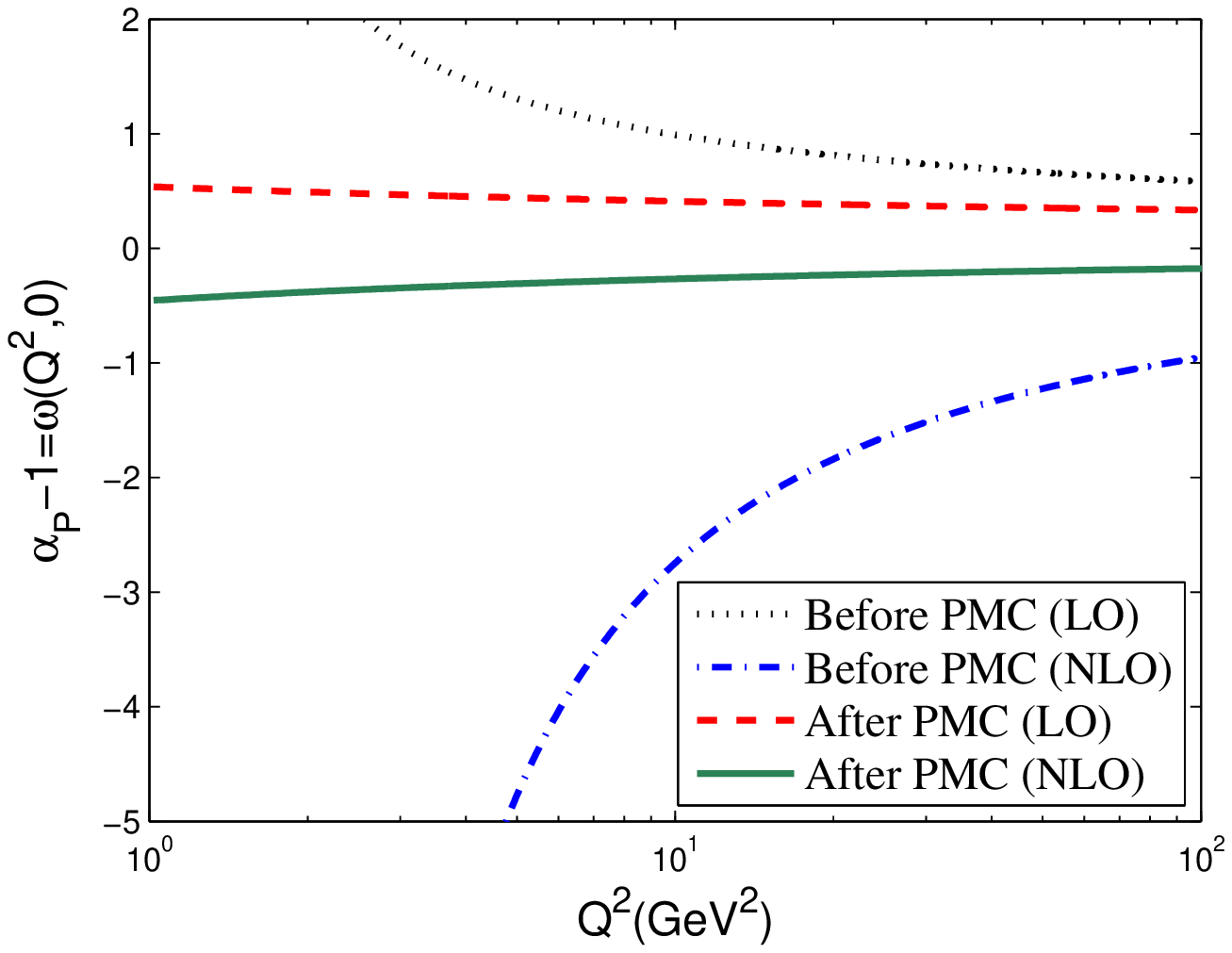}
\includegraphics[width=0.49\textwidth]{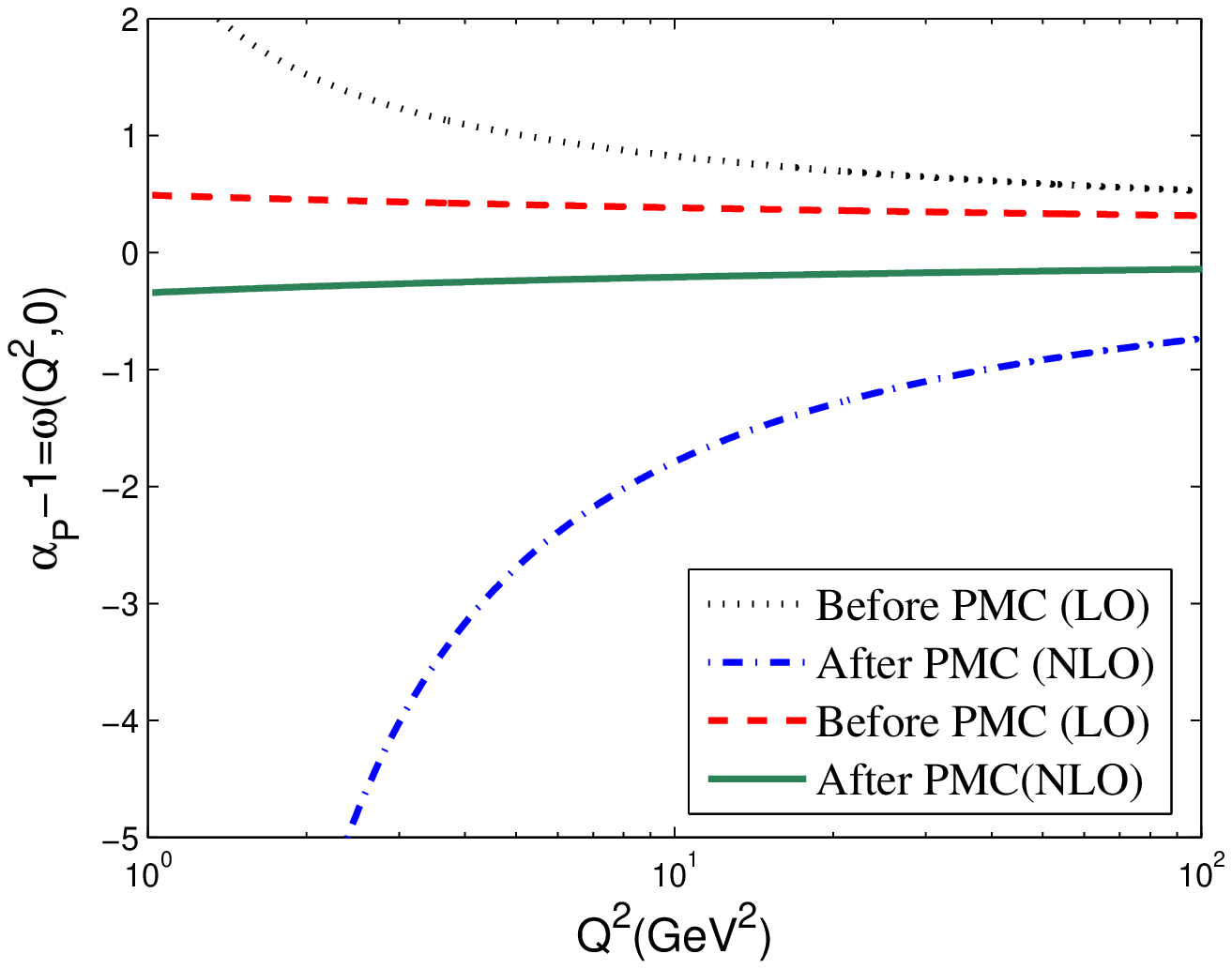}
\caption{The LO and NLO contributions of the BFKL Pomeron intercept versus $Q^2$ under the ${\rm MOM}$ scheme before or after PMC scale setting. The left diagram is for the Landau gauge with $\xi=0$ and the right one is for Fried-Yennie gauge with $\xi=3$.} \label{Plot:LONLO}
\end{figure}

\item We present the LO and NLO contributions of the BFKL Pomeron intercept before and after PMC scale setting in Fig.(\ref{Plot:LONLO}). Before PMC scale setting, the NLO term provides a larger negative contribution to the LO term, so one always obtains a negative Pomeron intercept which contradicts with the experimental result as mentioned in the Introduction. The PMC scale setting provides a systematic way to absorb the nonconformal $\{\beta_i\}$-terms into the running coupling, the divergent renormalon series is eliminated. Thus, in principle, the pQCD series becomes more convergent. It is found that after PMC scale setting, the magnitude of the NLO contribution becomes much smaller compared to that of conventional scale setting, which results in the required positive BFKL Pomeron intercept. In fact, because of the fact that the large log terms as $\ln(\mu_R/Q)$ shall always accompany with the $\{\beta_i\}$-terms, such large log terms are also eliminated after PMC scale setting. In this sense, the conventional idea of setting $\mu_R=Q$ is partly consistent with PMC scale setting, since some of the $\{\beta_i\}$-terms accompanying with the log terms are absorbed into the coupling constant through such naive choice.

\item The initial renormalization scale dependence, and hence the conventional scale dependence, can be greatly suppressed after PMC scale setting. In principle one needs to resum all known $\{\beta_i\}$-terms into PMC scales, and the initial renormalization scale independence may not be explicitly shown. For the present NLO correction, there is only one type of $\{\beta_i\}$-terms (i.e. the $\beta_0$-terms), thus to set the LO PMC scale is equivalent to resum all powers of $\beta_0$-terms into the PMC scale. In this sense, the large $\beta_0$-approximation suggested in the literature is consistent with our present treatment, e.g. for the quarkonium electromagnetic annihilation decays~\cite{chen}.

    As shown by Eq.(\ref{scale:MOM:PMC}), the resultant LO PMC scale and the $\omega_{\rm MOM}^{\rm PMC}(Q^{2},\nu)$ are independent of $\mu^{\rm init}_R$. Such initial scale independence shall be kept with high precision even after doing higher-order calculations, because of the fact that the higher-order $\{\beta_i\}$-terms shall be absorbed as the higher-order terms of the PMC scale itself and shall be further exponentially suppressed~\cite{PMC1}.

\begin{figure}[htb]
\centering
\includegraphics[width=0.49\textwidth]{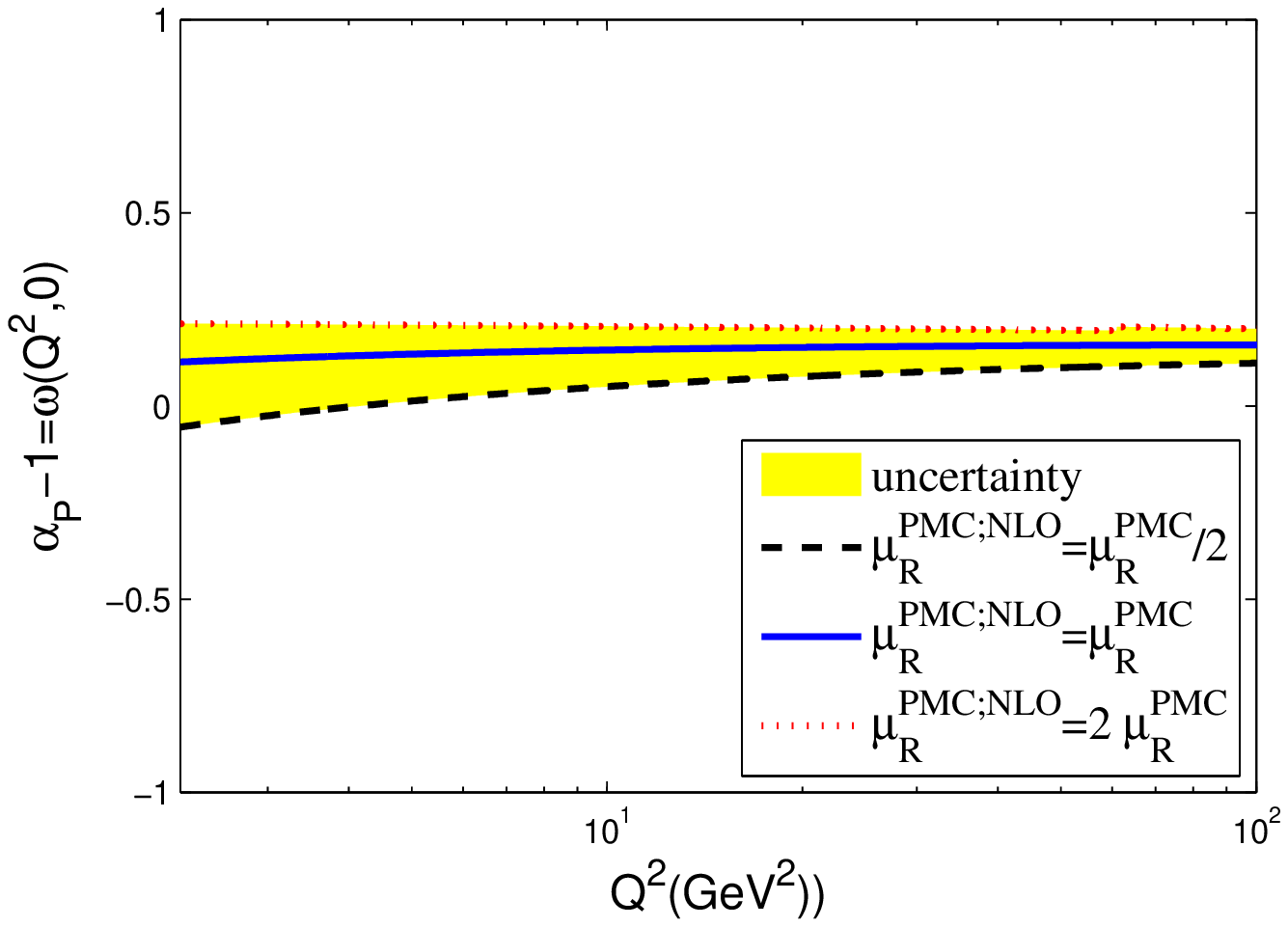}
\includegraphics[width=0.48\textwidth]{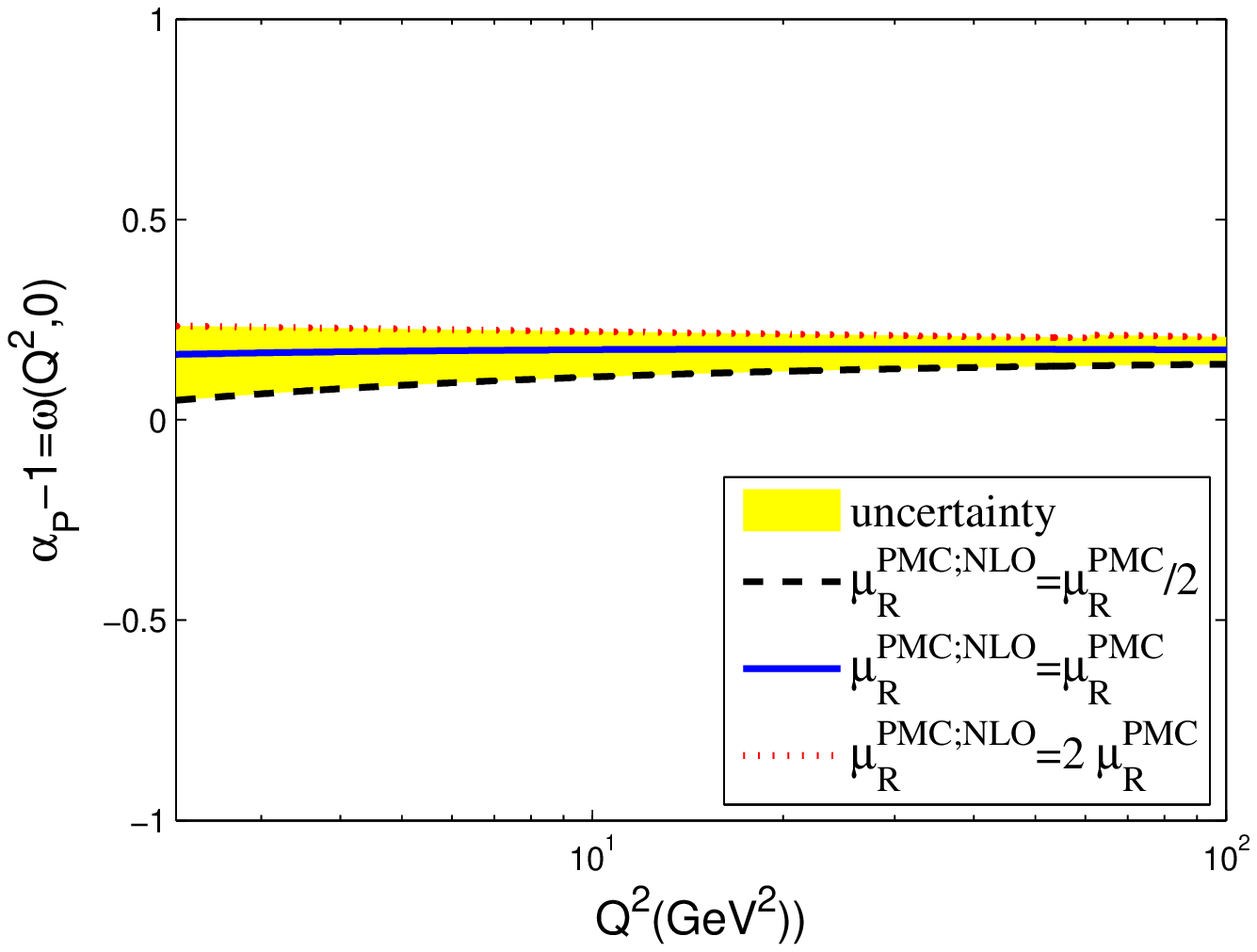}
\caption{A rough error analysis of the NLO BFKL Pomeron intercept $\omega(Q^{2},0)$ versus $Q^2$ under the ${\rm MOM}$ scheme and the PMC scale setting. The shaded band shows the residual renormalization scale uncertainty by varying the NLO PMC scale within the region of $[\mu^{\rm PMC;LO}_{R}/2, 2\mu^{\rm PMC;LO}_{R}]$ with $\mu^{\rm PMC;LO}_{R}=\mu^{\rm PMC}_{R}$ that is determined by Eq.(\ref{scale:MOM:PMC}). The left diagram is for the Landau gauge with $\xi=0$ and the right one is for the Fried-Yennie gauge with $\xi=3$. } \label{pmcdisscale}
\end{figure}

\item Because we do not have enough information to set the NLO PMC scale, so we have set it as $\mu^{\rm PMC;LO}_{R}$. The NLO-term is then the only term which contains a non-conformal contribution. In Fig.(\ref{pmcdisscale}), we present the NLO BFKL Pomeron intercept $\omega(Q^{2},0)$ versus $Q^2$ under the ${\rm MOM}$ scheme and the PMC scale setting. The shaded band shows the second type of the residual scale uncertainty by roughly varying the NLO PMC scale within the region of $\left[\mu^{\rm PMC;LO}_{R}/2, 2\mu^{\rm PMC;LO}_{R}\right]$ with $\mu^{\rm PMC;LO}_{R}=\mu^{\rm PMC}_{R}$ that is determined by Eq.(\ref{scale:MOM:PMC}). Through a comparison of Fig.(\ref{disscale}) and Fig.(\ref{pmcdisscale}), it is shown that the second type of residual scale dependence can also be greatly suppressed in comparison to the case of the conventional scale setting. This uncertainty, similar to the conventional scale setting, can give us some idea on how the unknown higher-order terms will affect the estimation, even though it only exposes the $\{\beta_i\}$-dependent non-conformal terms, not the entire perturbative series.

    Because the NLO terms provide large negative contributions to the Pomeron intercept even after PMC scale setting, the resultant second type of residual scale dependence is still relatively large. However, it is noted that such error estimation explicitly breaks the conformality of the pQCD series, and the present analysis of the scale uncertainty is conceptional, which can be greatly suppressed when we know the one order higher $\{\beta_i\}$-terms well. Hence, we shall always keep the choice of $\mu^{\rm PMC;NLO}_{R}=\mu^{\rm PMC;LO}_{R}$ in our following discussions.

\item Varying $Q^2$ within the region of $[1,100]\;{\rm GeV}^2$, we obtain $\omega_{\rm MOM}^{\rm PMC}(Q^{2},0)\in [0.082,0.158]$ for the Landau gauge of $\xi=0$, $\omega_{\rm MOM}^{\rm PMC}(Q^{2},0)\in [0.124,0.168]$ for the Feynman gauge of $\xi=1$, and $\omega_{\rm MOM}^{\rm PMC}(Q^{2},0)\in [0.149,0.176]$ for the Fried-Yennie gauge of $\xi=3$, respectively.

\begin{table}[htb]
\centering
\begin{tabular}{|c||c|c|c||c|c|c|}
\hline
 & \multicolumn{3}{c||}{$\xi=0$}  &  \multicolumn{3}{c|}{$\xi=3$}  \\
\hline
$Q^2$ & $1~{\rm GeV}^2$ & $15~{\rm GeV}^2$ & $100~{\rm GeV}^2$  & $1~{\rm GeV}^2$ & $15~{\rm GeV}^2$ & $100~{\rm GeV}^2$ \\
\hline
$\omega_{\rm MOM}^{\rm PMC}(Q^{2},0)$ & 0.082 & 0.149 & 0.158 & 0.149 & 0.176 & 0.175 \\
\hline
\end{tabular}
\caption{The NLO BFKL Pomeron intercept $\omega_{\rm MOM}^{\rm PMC}(Q^{2},0)$ after PMC scale setting for three typical virtualities of the reggeized gluon. The results for the Landau gauge with $\xi=0$ and the Fried-Yennie gauge with $\xi=3$ are presented respectively.} \label{t:initialscale}
\end{table}

    We present the NLO BFKL Pomeron intercept $\omega_{\rm MOM}^{\rm PMC}(Q^{2},0)$ at $Q^2=1 {\rm GeV}^2$, $15 {\rm GeV}^2$ and $100 {\rm GeV}^2$ in Table \ref{t:initialscale}, where the results for the Landau gauge and the Fried-Yennie gauge are presented respectively. Such weak $Q^2$-dependence of the Pomeron intercept agrees with the previous statements drawn in Ref.\cite{add1}, that is, the physical results should not depend on $Q^2$ in the Regge factors because its change is compensated by the corresponding modification of the impact factors and the kernel, and hence it does not have any influence on the correction to the Pomeron intercept.

\begin{figure}[htb]
\centering
\includegraphics[width=0.60\textwidth]{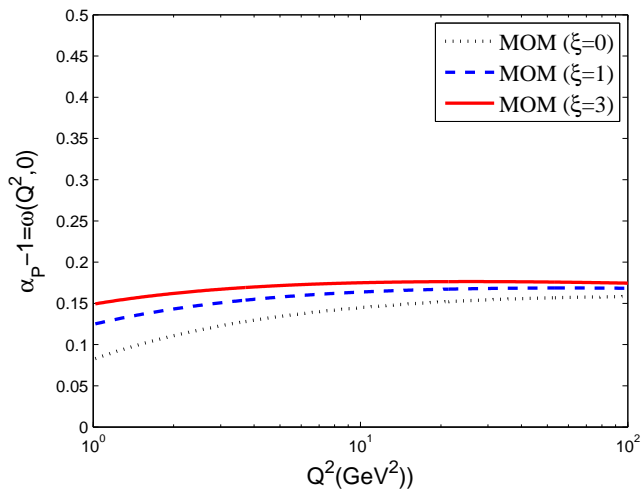}
\caption{The NLO Pomeron intercept versus $Q^2$ under the MOM scheme and three typical gauges, where the dotted, the dashed and the solid lines are for $\xi=0$, $\xi=1$ and $\xi=3$ respectively. } \label{Plot:afterPMC}
\end{figure}

    As shown by Fig.(\ref{disscale}), the gauge dependence is the weak point of the MOM scheme itself~\cite{MOM}. However, in comparison to the case of conventional scale setting, the Pomeron intercept after PMC scale setting has a much weaker dependence on the reggeized gluon virtuality $Q^2$, which is a good property when applies to higher-energy phenomenology. One example of which will be shown in the following subsection. More explicitly, for the case of $Q^2=15\;{\rm GeV}^2$, we obtain $\omega_{\rm MOM}^{\rm PMC}(Q^{2},0) =0.166^{+0.010}_{-0.017}$, where the gauge dependence is about $10\%$ with the central value for $\xi=1$, the upper error for $\xi=3$ and the lower error for $\xi=0$, respectively. Fig.(\ref{Plot:afterPMC}) shows this point more clearly, which shows the behavior of the NLO Pomeron intercept versus $Q^2$ for the three different gauges.

\end{enumerate}

\subsection{A discussion on the photon-photon collision experiment}
\label{sec:photonphoton}

We choose the scattering of virtual photons as an example to compare our PMC calculations with the experiment. The experimental analysis of this process can be useful to constrain the QCD dynamics. The BFKL Pomeron theory is applicable for this process, since the photon virtuality provides the hard scale to make pQCD applicable. In fact, the photon-photon collision plays a special role in QCD, i.e. it provides a good opportunity to test the high-energy asymptotic behavior of QCD~\cite{gamma0,gammagamma,gammagammaPMS,gamma2,gamma5,gamma6}.

\begin{figure}[htb]
\centering
\includegraphics[width=0.60\textwidth]{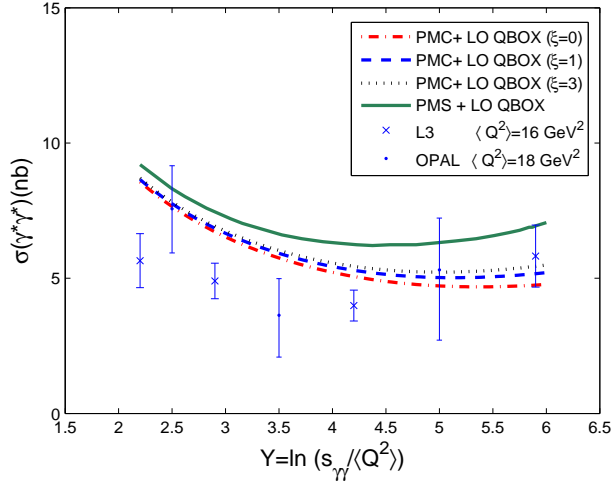}
\caption{The energy dependence of the total cross section for the highly virtual photon-photon collisions with $Q^2=17\;{\rm GeV}^2$ predicted by NLO BFKL under PMC and PMS scale setting compared with OPAL~\cite{OPAL} ($\langle Q^2\rangle =18 GeV^2$) and L3~\cite{L3} ($\langle Q^2\rangle =16\; {\rm GeV}^2$) data from the previous LEP-2 experiment at CERN. } \label{Plot:gammagamma}
\end{figure}

In Fig.(\ref{Plot:gammagamma}), we show the energy dependence of the total cross section for the highly virtual photon-photon collisions with $Q^2=17\;{\rm GeV}^2$ predicted by NLO BFKL under PMC and PMS scale setting. In the estimation, similar to Ref.\cite{gammagamma}, we have included the contributions from the LO quark box diagrams~\cite{LOQBOX} and NLO BFKL resummation. In Fig.(\ref{Plot:gammagamma}), we present the OPAL~\cite{OPAL} ($\langle Q^2\rangle =18\;{\rm GeV}^2$) and the L3~\cite{L3} ($\langle Q^2\rangle =16\;{\rm GeV}^2$) data from the previous LEP-2 experiment with the collision energy $E_{cm}\in(189,209)$ GeV. As a comparison, we also present the theoretical calculation from the principle of minimum sensitivity (PMS) scale setting~\cite{gammagammaPMS}, which can also give reasonable estimation as the case of PMC scale setting and shows a large bias from the data in lower $Q^2$ region. It shows that after PMC scale setting, the gauge choices under the MOM scheme is very small, especially in high collision energy region. One may observe that the theoretical results from our calculation and the PMS method are larger than the data in lower energy region. This is due to that at present, we only take the LO impact factor of the virtual photon into consideration, while it is found there is a large negative contribution from the NLO correction to the impact factor in lower energy region~\cite{NLOQBOX}.

\begin{figure}
\centering
\includegraphics[width=0.60\textwidth]{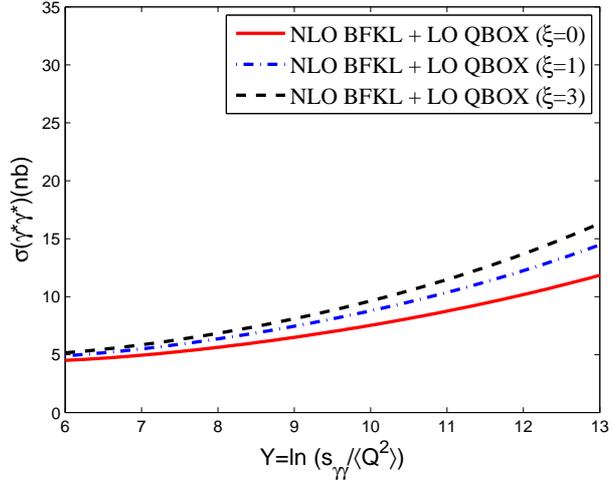}
\caption{The energy dependence of the total cross section of highly virtual photon-photon collisions with $\langle Q^2\rangle\equiv 20\; {\rm GeV}^2$ predicted by NLO BFKL under PMC scale setting for future linear colliders with the collision energy up to 2 TeV~\cite{ILC}. } \label{Plot:gammagammaILC}
\end{figure}

As a final remark. In Fig.(\ref{Plot:gammagammaILC}), we present the energy dependence of the total cross section for $\langle Q^2\rangle=20\;{\rm GeV}^2$ in the energy region of future linear colliders with collision energy up to 2 TeV~\cite{ILC}. Three curves in Fig.(\ref{Plot:gammagammaILC}) show the results from three gauge choices under the MOM scheme, respectively. When the collision energy $\sqrt{s_{\gamma\gamma}}=1\;{\rm TeV}$, we obtain
\begin{eqnarray}
\sigma(\gamma^*\gamma^*) &=& 8.51\;{\rm nb}\;\;\;\;\;  {\rm for}\; \xi=0 , \\ \sigma(\gamma^*\gamma^*) &=& 10.06\;{\rm nb}\;\;\; {\rm for}\; \xi=1 , \\ \sigma(\gamma^*\gamma^*) &=& 11.12\;{\rm nb}\;\;\; {\rm for}\; \xi=3 .
\end{eqnarray}
When the collision energy $\sqrt{s_{\gamma\gamma}}=2\;{\rm TeV}$, we obtain
\begin{eqnarray}
\sigma(\gamma^*\gamma^*) &=& 10.50\;{\rm nb}\;\;\; {\rm for}\; \xi=0 ,\\ \sigma(\gamma^*\gamma^*) &=& 12.67\;{\rm nb}\;\;\; {\rm for}\; \xi=1 ,\\ \sigma(\gamma^*\gamma^*) &=& 14.19\;{\rm nb}\;\;\; {\rm for}\; \xi=3 .
\end{eqnarray}
This shows a measurement of this cross section at the future linear colliders with large collision energy and high luminosity shall provide an excellent test of the BFKL Pomeron.

\section{Summary} \label{sec:4}

In the paper, we have studied the BFKL Pomeron intercept up to NLO level by using PMC scale setting. It is shown that in different to the conventional scale setting, we can obtain a reasonable intercept by using PMC scale setting.

In doing PMC scale setting, we first transform the Pomeron intercept under the $\overline{\rm MS}$-scheme to the one under the physical MOM-scheme. Then, after applying PMC scale setting, we obtain $\omega_{\rm MOM}^{\rm PMC}(Q^{2},0)\in [0.082,0.158]$ for the Landau gauge, $\omega_{\rm MOM}^{\rm PMC}(Q^{2},0)\in [0.124,0.168]$ for the Feynman gauge, and $\omega_{\rm MOM}^{\rm PMC}(Q^{2},0)\in [0.149,0.176]$ for the Fried-Yennie gauge, where $Q^2\in[1,100]\;{\rm GeV}^2$. Using the BFKL Pomeron intercept as an explicit example, we have shown that there are several good features after PMC scale setting:
\begin{itemize}
\item The unreasonable negative estimation of the Pomeron intercept and the large scale uncertainty under conventional scale setting can be greatly cured. The LO PMC scale and hence the NLO BFKL Pomeron intercept is highly independent of the initial renormalization scale. This is because the only $\{\beta_i\}$-terms (i.e. the $\beta_0$-terms) at the NLO level rightly determines the physical behavior of the LO coupling constant. There are residual PMC scale dependence, which comes form the unknown $\{\beta_i\}$-terms at the NNLO level or higher. It is shown that such residual scale dependence can be highly suppressed.

\item As shown by Fig.(\ref{Plot:afterPMC}), the NLO BFKL Pomeron intercept has a weak dependence on the virtuality of the reggeized gluon, and also has a small gauge dependence for the MOM-scheme. These are good properties to apply to the high-energy phenomenology. For example, we have applied the BFKL Pomeron intercept to the photon-photon collision process, and compared the theoretical predictions with the data from the OPAL and L3 experiments. In Fig.(\ref{Plot:gammagammaILC}), we present the energy dependence of the total cross section for the photon-photon collision at $\langle Q^2\rangle=20\;{\rm GeV}^2$ with its collision energy up to 2 TeV. We expect a verification of the BFKL Pomeron at the future International Linear Collider.

\item The PMC scale setting provides a systematic way to absorb the nonconformal $\{\beta_i\}$-terms into the running coupling, the divergent renormalon series $(n! \alpha_s^n \beta_i^n)$ is eliminated. Thus, as shown clearly in Fig.(\ref{Plot:LONLO}), the pQCD series becomes more convergent and we obtain a more credible estimation of Pomeron intercept.
\end{itemize}

\hspace{1cm}

{\bf\Large Acknowledgement:} The authors would like to thank Stanley Brodsky and Matin Mojaza for helpful discussions. This work was supported in part by the Fundamental Research Funds for the Central Universities under Grant No.WLYJSBJRCGR201106 and No.CQDXWL-2012-Z002, by Natural Science Foundation of China under Grant No.11075225 and No.11275280, and by the Program for New Century Excellent Talents in University under Grant NO.NCET-10-0882.

\end{document}